\documentclass[10pt]{article}
\usepackage{amsmath,amssymb}
\usepackage[dvips]{epsfig}
\usepackage{graphicx}

{\hspace*{\fill}$\rule{.3\baselineskip}{.35\baselineskip}$\end{trivlist}}

\renewcommand{\phi}{\varphi}
\newcommand{\be}{\begin{eqnarray}}
\newcommand{\ee}{\end{eqnarray}}

\begin{document}

\title{\bf New integrable semi-discretizations \\ of the coupled nonlinear Schr\"{o}dinger equations}

\author{Sylvie A. Bronsard and Dmitry E. Pelinovsky \\
{\small Department of Mathematics and Statistics, McMaster
University, Hamilton, Ontario, Canada, L8S 4K1} }

\date{\today}
\maketitle

\begin{abstract}
We have undertaken an algorithmic search for new integrable semi-discretizations of physically 
relevant nonlinear partial differential equations. The search is performed 
by using a compatibility condition for the discrete Lax operators and symbolic computations.
We have discovered a new integrable system of coupled nonlinear Schr\"{o}dinger equations
which combines elements of the Ablowitz--Ladik lattice and the triangular--lattice ribbon studied by 
Vakhnenko. We show that the continuum limit of the new integrable system is given by
uncoupled complex modified Korteweg--de Vries equations and
uncoupled nonlinear Schr\"{o}dinger equations.
\end{abstract}

\textbf{Keywords:} integrable semi-discretizations, derivative nonlinear Schr\"{o}dinger equation,
massive Thirring model, Kaup--Newell spectral problem, Ablowitz--Ladik lattice.

\section{Introduction}

As was understood long ago, many nonlinear partial differential equations 
integrable with the inverse scattering transform can be semi-discretized in spatial coordinates 
in such a way as to preserve integrability.
The pioneer example is the Ablowitz--Ladik lattice \cite{AL,AL2}, an integrable 
semi-discretization of the integrable nonlinear Schr\"{o}dinger equation. The Ablowitz--Ladik lattice has inspired many
groups to search for integrable semi-discretizations of other nonlinear evolution equations, e.g. \cite{Trubach,Gerjikov,Kulish,Ts1,Ts2}. 
The nonlinear ladder equation,
the Toda lattice, the discrete modified Korteweg--de Vries equation, the discrete sine--Gordon equation in characteristic coordinates,
and the nonlinear self-dual network equations are examples of integrable semi-discrete evolution equations
related to the Ablowitz--Kaup--Newel--Segur (AKNS) spectral problem \cite{AC}.

Other spectral problems have been semi-discretized only very recently. Tsuchida \cite{Tsuchida} considered
nonlinear evolution equations related to the Kaup--Newell spectral problem and constructed integrable semi-discretizations
of the derivative nonlinear Schr\"{o}dinger equation, the Chen--Lee-Liu equation, and the Gerdjikov--Ivanov equations.
The coupled Yajima--Oikawa system was semi-discretized by using the Hirota bilinear method in \cite{ChenChen}.
Generalization of integrable discretizations in the space of two spatial dimensions was considered by
Zakharov \cite{Zaharov} by using an algebro-geometric approach. The integrable triangular--lattice ribbon
was recently studied by Vakhnenko \cite{V1,V2} (see also a review in \cite{Vakhnenko})
who further generalized the discrete AKNS spectral problem
by including quadratic dependence on the spectral parameter.

In a similar vein, one motivation for our work is to find an integrable
semi-discretization of the massive Thirring model (MTM) \cite{KN,KM},
\begin{equation}
\label{MTM}
\left\{ \begin{array}{l}
i(u_t+u_x)+v+u|v|^2=0, \\
i(v_t-v_x)+u+v|u|^2=0, \end{array} \right.
\end{equation}
which has been used very recently in many studies related to stability of one-dimensional Dirac solitons
\cite{Comech,Yusuke1,Mertens,Yusuke2}. Numerical methods based on various spatial semi-discretizations of the MTM
were found to suffer from numerical instabilities and artifacts \cite{Kevrekidis,Shao}. If we find a semi-discretization which
preserves the integrability scheme of the MTM, then the discrete MTM should model stable
Dirac solitons without numerical artifacts.

The integrability scheme for the MTM is related to the Kaup--Newell spectral problem \cite{KN78}. 
The same spectral problem is also related to the integrability scheme for the derivative nonlinear Schr\"{o}dinger (dNLS) equation,
\begin{equation}
\label{dnls}
i u_t + u_{xx} + i (|u|^2 u)_x = 0.
\end{equation}
In more details, the dNLS equation (\ref{dnls}) is the compatibility condition $\vec{\phi}_{xt}=\vec{\phi}_{tx}$
for the system of linear equations
\begin{equation} \label{laxeq-dnls}
\vec{\phi}_x = L(\lambda;u) \vec{\phi}\quad  \mbox{and}\quad \vec{\phi}_t = A(\lambda;u) \vec{\phi},
\end{equation}
where $\lambda$ is a spectral parameter, while $L(\lambda;u)$ and $A(\lambda;u)$ are matrix operators given by
\begin{equation} \label{lablax1-dnls}
L(\lambda;u) = - i \lambda^2 \sigma_3 + \lambda \left(\begin{matrix} 0 & u \\ -\bar{u} & 0 \end{matrix}\right)
\end{equation}
and
\begin{equation} \label{lablax2-dnls}
A(\lambda;u) = i (\lambda^2 |u|^2 - 2\lambda^4) \sigma_3 + \lambda
\left(\begin{matrix} 0 & 2 \lambda^2 u - |u|^2 u \\ -2 \lambda^2 \bar{u} + |u|^2 \bar{u} & 0 \end{matrix}\right)
+ i \lambda \left(\begin{matrix} 0 & u_x \\ \bar{u}_x & 0 \end{matrix}\right),
\end{equation}
where $\sigma_3 = {\rm diag}(1,-1)$ is Pauli's matrix.

Similarly, the MTM system (\ref{MTM}) is the compatibility condition $\vec{\phi}_{xt}=\vec{\phi}_{tx}$
for the system of linear equations
\begin{equation} \label{laxeq}
\vec{\phi}_x = L(\lambda;u,v) \vec{\phi}\quad  \mbox{and}\quad \vec{\phi}_t = A(\lambda;u,v) \vec{\phi},
\end{equation}
where $\lambda$ is a spectral parameter, while $L(\lambda;u,v)$ and $A(\lambda;u,v)$ are matrix operators given by
\begin{equation} \label{lablax1}
L(\lambda;u,v) = \frac{i}{4}(|u|^2-|v|^2)\sigma_3-\frac{i\lambda}{2}\left(\begin{matrix} 0 & \overline{v} \\ v & 0 \end{matrix}\right)
+ \frac{i}{2\lambda} \left(\begin{matrix} 0 & \overline{u} \\ u & 0 \end{matrix}\right) +\frac{i}{4}\left(\lambda^2-\frac{1}{\lambda^2}\right)\sigma_3
\end{equation}
and
\begin{equation} \label{lablax2}
A(\lambda;u,v) = -\frac{i}{4}(|u|^2+|v|^2)\sigma_3-\frac{i\lambda}{2}\left(\begin{matrix} 0 & \overline{v} \\ v & 0 \end{matrix}\right) - \frac{i}{2\lambda} \left(\begin{matrix} 0 & \overline{u} \\ u & 0 \end{matrix}\right) +\frac{i}{4}\left(\lambda^2+\frac{1}{\lambda^2}\right)\sigma_3.
\end{equation}
Compared to the matrix operators in (\ref{lablax1-dnls})--(\ref{lablax2-dnls}),
both $L$ and $A$ in (\ref{lablax1})--(\ref{lablax2}) depend quadratically on $\lambda$ and $\lambda^{-1}$, which makes
analysis of the inverse scattering transform for the MTM sufficiently difficult \cite{Villarroel}. By performing
a transformation of the physical coordinates $x$ and $t$ to the characteristic coordinates $\xi = x-t$ and $\eta = x+t$, 
one can rewrite the MTM system (\ref{MTM})
and the Lax pair (\ref{laxeq}) in the form associated with the Kaup--Newell operator $L$ in (\ref{lablax1-dnls}).
However, this transformation changes the Cauchy problem for the MTM system (\ref{MTM})
to the Goursat problem in characteristic coordinates and vice versa.

Tsuchida \cite{Tsuchida} obtained semi-discretizations of the dNLS equation (\ref{dnls}) and the MTM in characteristic coordinates
by using the gauge transformation of the Kaup-Newell spectral problem
to the AKNS spectral problem and by searching for a generalized spatial discretization of the AKNS problem.
However, as is explained above, these semi-discretizations are not useful in the context of the Cauchy problem for the MTM system
in physical coordinates (\ref{MTM}).

We have undertaken here a systematic search for the class of semi-discrete matrices $L$ and $A$ with a polynomial
dependence on $z$ and $z^{-1}$ up to the quadratic (for $L$) and quartic (for $A$) orders,
where $z$ is a new spectral parameter. As an outcome of our algorithmic computations,
we have obtained a new semi-discretization of the coupled nonlinear Schr\"{o}dinger equations. This new semi-discretization
coincides with the higher-order commuting flow of the triangular--lattice ribbon \cite{V1,V2}.

The rest of this paper is organized as follows. Section 2 presents the discrete spectral problems for
the Ablowitz--Ladik lattice, triangular--lattice ribbon, and the newly derived semi-discretization
of the coupled nonlinear Schr\"{o}dinger equations. Section 3 contains a study of the continuum limit
in the new semi-discrete system. Section 4 concludes the paper with a summary.

\section{Semi-discretizations of the nonlinear Schr\"{o}dinger equations}

We are looking for an integrable semi-discrete system which appears as
a compatibility condition for the system of linear equations
\begin{equation} \label{LAX}
\vec{\phi}_{n+1} = L_n(z) \vec{\phi}_n \quad  \mbox{and}\quad \frac{d}{dt} \vec{\phi}_n = A_n(z) \vec{\phi}_n,
\end{equation}
where $z$ is a spectral parameter, $n \in \mathbb{Z}$, $t \in \mathbb{R}$, while
$L_n(z)$ and $A_n(z)$ are matrix operators containing potentials satisfying
the compatibility condition
\begin{equation}
\label{compatibility}
\frac{d}{dt} L_n(z) = A_{n+1}(z) L_n(z) - L_n(z) A_n(z).
\end{equation}

{\em The Ablowitz--Ladik lattice} derived in \cite{AL,AL2} corresponds to the choice
\begin{equation} \label{ALlax}
L_n(z) = \left[ \begin{array}{cc} z & q_n \\ -\bar{q}_n & z^{-1} \end{array} \right], \quad
A_n(z) = \left[ \begin{array}{cc} a z^2 + a q_n \bar{q}_{n-1} & a q_n z + \bar{a} q_{n-1} z^{-1} \\
- a \bar{q}_{n-1} z - \bar{a} \bar{q}_n z^{-1} & \bar{a} z^{-2} + \bar{a} q_{n-1} \bar{q}_n \end{array} \right],
\end{equation}
where $a \in \mathbb{C}$ is arbitrary parameter
and the complex-conjugate symmetry is preserved for the complex-valued potential $\{ q_n \}_{n \in \mathbb{Z}}$.
Substituting (\ref{ALlax}) into (\ref{compatibility}) yields the Ablowitz--Ladik lattice
\begin{equation}
\label{ALlattice}
\frac{d q_n}{dt} = \alpha (q_{n+1}-q_{n-1})(1 + |q_n|^2) + i \beta (q_{n+1}+q_{n-1}) (1 + |q_n|^2),
\end{equation}
where we have used $a = \alpha + i \beta$ with $\alpha,\beta \in \mathbb{R}$. The $\alpha$ part
of this system is also referred to as the discrete modified Korteweg-de Vries equation,
while the $\beta$ part is referred to as the discrete nonlinear Schr\"{o}dinger equation \cite{AC}.
The two parts are related by the staggering transformation
$$
q_n \mapsto i^n q_n, \quad n \in \mathbb{Z}.
$$

{\em The triangular--lattice ribbon} derived in \cite{V1,V2} corresponds to the choice
\begin{equation} \label{Vakhlax1}
L_n(z) = \left[ \begin{array}{cc} z^2 -\bar{q}_n r_n & q_n z + r_n z^{-1} \\
-\bar{r}_n z - \bar{q}_n z^{-1} & z^{-2} - q_n \bar{r}_n \end{array} \right]
\end{equation}
and
\begin{equation} \label{Vakhlax2}
A_n(z) = \left[ \begin{array}{cc} a z^2 + a q_n \bar{r}_{n-1} & a q_n z + \bar{a} r_{n-1} z^{-1} \\
-a \bar{r}_{n-1} z - \bar{a} \bar{q}_n z^{-1} & \bar{a} z^{-2} + \bar{a} r_{n-1} \bar{q}_n \end{array} \right],
\end{equation}
where $a \in \mathbb{C}$ is arbitrary parameter and the complex-conjugate symmetry
is preserved for the complex-valued potentials $\{ q_n, r_n \}_{n \in \mathbb{Z}}$.
Substituting (\ref{Vakhlax1}) and (\ref{Vakhlax2}) into (\ref{compatibility}) yields
the triangular--lattice ribbon:
\begin{eqnarray}
\label{Vakh1}
\frac{d q_n}{dt} & = & \alpha  (r_n - r_{n-1}) (1 + |q_n|^2)  + i \beta (r_n + r_{n-1}) (1 + |q_n|^2), \\
\label{Vakh2}
\frac{d r_n}{dt} & = & \alpha  (q_{n+1} - q_n) (1 + |r_n|^2)  + i \beta (q_{n+1} + q_n) (1 + |r_n|^2),
\end{eqnarray}
where we have used $a = \alpha + i \beta$ with $\alpha,\beta \in \mathbb{R}$. The $\alpha$ part
of this system is referred to as the nonlinear self-dual network equations \cite{AC}.
The $\beta$ part can be transformed to the $\alpha$ part by
the staggering transformation
$$
q_n \mapsto (-1)^n q_n, \quad r_n \mapsto -i (-1)^n r_n, \quad n \in \mathbb{Z}.
$$

Our search of the matrix operators $L_n(z)$ and $A_n(z)$ satisfying the compatibility condition (\ref{compatibility})
generalizes the choices (\ref{ALlax}) and (\ref{Vakhlax1})--(\ref{Vakhlax2}).
We have considered a general quadratic polynomial in $z$ and $z^{-1}$ for $L_n(z)$
and a general quartic polynomial in $z$ and $z^{-1}$ for $A_n(z)$. By working with
the symbolic computation software based on Wolfram's MATHEMATICA, we were able to
satisfy the compatibility condition (\ref{compatibility}) in each order of $z$ and $z^{-1}$
if the matrix operators $L_n(z)$ and $A_n(z)$ are given in the form
\begin{equation} \label{SylvLax}
L_n(z) = \left[ \begin{array}{cc} z^2 - \bar{q}_n r_n & q_n z + r_n z^{-1} \\
-\bar{r}_n z - \bar{q}_n z^{-1} & z^{-2} - q_n \bar{r}_n \end{array} \right],
\quad A_n(z) = \left[ \begin{array}{cc} A_{11}(z) & A_{12}(z) \\ A_{21}(z) & A_{22}(z) \end{array} \right]
\end{equation}
with
\begin{eqnarray*}
A_{11}(z) & = & a  z^4 + a q_n \bar{r}_{n-1} z^2 + a \left( q_n \bar{q}_{n-1} + r_n \bar{r}_{n-1}
+ q_n \bar{q}_{n-1} |r_{n-1}|^2 + |q_n|^2 r_n \bar{r}_{n-1} + q_n^2 \bar{r}_{n-1}^2 \right) \\
& \phantom{t} & \phantom{texttext} - \bar{a} \bar{q}_n r_{n-1} z^{-2}, \\
A_{12}(z) & = & a q_n z^3 + a (r_n + |q_n|^2 r_n - q_n^2 \bar{r}_{n-1}) z
+ \bar{a} ( q_{n-1} + q_{n-1} |r_{n-1}|^2 + \bar{q}_n r_{n-1}^2) z^{-1} + \bar{a} r_{n-1} z^{-3}, \\
A_{21}(z) & = & -a \bar{r}_{n-1} z^3 - a (\bar{q}_{n-1} + \bar{q}_{n-1} |r_{n-1}|^2 + q_n \bar{r}_{n-1}^2) z
- \bar{a} ( \bar{r}_n + |q_n|^2 \bar{r}_{n} + \bar{q}^2_n r_{n-1}) z^{-1} - \bar{a} \bar{q}_{n} z^{-3}, \\
A_{22}(z) & = & - a q_n \bar{r}_{n-1} z^2 + \bar{a} \left( q_{n-1} \bar{q}_n + r_{n-1} \bar{r}_n + q_{n-1} \bar{q}_n |r_{n-1}|^2
+ |q_n|^2 r_{n-1} \bar{r}_n + \bar{q}_n^2 r_{n-1}^2 \right) \\
& \phantom{t} & \phantom{texttext} + \bar{a} \bar{q}_n r_{n-1} z^{-2} + \bar{a} z^{-4}.
\end{eqnarray*}
The two potentials $\{ q_n, r_n \}_{n \in \mathbb{Z}}$ satisfy the lattice differential equations
in the form:
\begin{eqnarray}
\nonumber
\frac{d q_n}{dt} & = & \left[ a q_{n+1} (1 + |r_n|^2) - \bar{a} q_{n-1} (1 + |r_{n-1}|^2)
+ \bar{q}_n (a r_n^2 - \bar{a} r_{n-1}^2) \right. \\
\label{Sylvlattice1} & \phantom{t} & \left. + q_n (a r_n \bar{r}_{n-1}-\bar{a} r_{n-1} \bar{r}_n) \right] (1 + |q_n|^2), \\
\nonumber
\frac{d r_n}{dt} & = & \left[ a r_{n+1} (1 + |q_{n+1}|^2) - \bar{a} r_{n-1} (1 + |q_n|^2)
+ \bar{r}_n (a q_{n+1}^2-\bar{a} q_{n}^2) \right. \\
\label{Sylvlattice2} & \phantom{t} & \left. + r_n (a q_{n+1} \bar{q}_{n}- \bar{a} q_{n} \bar{q}_{n+1}) \right] (1 + |r_n|^2).
\end{eqnarray}
Comparing the matrix operators $L_n(z)$ in (\ref{Vakhlax1}) and (\ref{SylvLax}), we can see that
they are identical to each other. On the other hand,
the matrix operator $A_n(z)$ in (\ref{SylvLax}) yields the next commuting flow
to the matrix operator $A_n(z)$ in (\ref{Vakhlax2}). Hence,
the new system of lattice differential equations (\ref{Sylvlattice1})--(\ref{Sylvlattice2}) is
the next commuting flow of the triangular--lattice ribbon (\ref{Vakh1})--(\ref{Vakh2}).

\section{Continuum limit of the semi-discrete equations}

Here we derive the continuum limit of the semi-discrete equations (\ref{Sylvlattice1})--(\ref{Sylvlattice2})
and compare them with integrable continuous nonlinear equations.
Setting $a = 1$ in (\ref{Sylvlattice1})--(\ref{Sylvlattice2}) yields
\begin{eqnarray}
\nonumber
\frac{d q_n}{dt} & = & \left[ q_{n+1} (1 + |r_n|^2)-q_{n-1} (1 + |r_{n-1}|^2)
+ \bar{q}_n (r_n^2-r_{n-1}^2) \right. \\
\label{Sylvlattice5} & \phantom{t} & \left. + q_n (r_n \bar{r}_{n-1}-r_{n-1} \bar{r}_n) \right] (1 + |q_n|^2), \\
\nonumber
\frac{d r_n}{dt} & = & \left[ r_{n+1} (1 + |q_{n+1}|^2) - r_{n-1} (1 + |q_n|^2)
+ \bar{r}_n (q_{n+1}^2-q_{n}^2) \right. \\
\label{Sylvlattice5} & \phantom{t} & \left. + r_n (q_{n+1} \bar{q}_{n}-q_{n} \bar{q}_{n+1}) \right] (1 + |r_n|^2).
\end{eqnarray}
By taking the asymptotic ansatz
$$
\left\{ \begin{array}{l}
q_n(t) = \epsilon Q(\epsilon (n+2t), \epsilon^3 t) + \mathcal{O}(\epsilon^3), \\
r_n(t) = \epsilon R(\epsilon (n+2t), \epsilon^3 t) + \mathcal{O}(\epsilon^3),
\end{array} \right.
$$
we obtain the system of coupled complex modified Korteweg--de Vries equations
at the leading order of $\mathcal{O}(\epsilon^4)$:
\begin{eqnarray}
\label{KdV1}
Q_{\tau} = \frac{1}{3} Q_{\xi \xi \xi} + 2 (|Q|^2+|R|^2) Q_{\xi} + 2 (Q \bar{R} + \bar{Q} R) R_{\xi}, \\
\label{KdV2}
R_{\tau} = \frac{1}{3} R_{\xi \xi \xi} + 2 (|Q|^2+|R|^2) R_{\xi} + 2 (Q \bar{R} + \bar{Q} R) Q_{\xi},
\end{eqnarray}
where $\xi = \epsilon (n+2t)$ and $\tau = \epsilon^3 t$. Although the system of coupled equations
(\ref{KdV1})--(\ref{KdV2}) may look as a new integrable system, it has a simple reduction
to uncoupled complex modified Korteweg-de Vries equations. Indeed, let $U := Q + R$ and
$V := Q - R$. Then, adding and subtracting (\ref{KdV1}) and (\ref{KdV2}) yield
the following two uncoupled complex modified Korteweg-de Vries equations:
\begin{eqnarray}
\label{KdV3}
U_{\tau} = \frac{1}{3} U_{\xi \xi \xi} + 2 |U|^2 U_{\xi}, \\
\label{KdV4}
V_{\tau} = \frac{1}{3} V_{\xi \xi \xi} + 2 |V|^2 V_{\xi}.
\end{eqnarray}
Setting $a = i$ in (\ref{Sylvlattice1})--(\ref{Sylvlattice2}) yields
\begin{eqnarray}
\nonumber
\frac{d q_n}{dt} & = & i \left[ q_{n+1} (1 + |r_n|^2) + q_{n-1} (1 + |r_{n-1}|^2)
+ \bar{q}_n (r_n^2+r_{n-1}^2) \right. \\
\label{Sylvlattice3} & \phantom{t} & \left. + q_n (r_n \bar{r}_{n-1}+r_{n-1} \bar{r}_n) \right] (1 + |q_n|^2), \\
\nonumber
\frac{d r_n}{dt} & = & i\left[ r_{n+1} (1 + |q_{n+1}|^2) + r_{n-1} (1 + |q_n|^2)
+ \bar{r}_n (q_{n+1}^2 + q_{n}^2) \right. \\
\label{Sylvlattice4} & \phantom{t} & \left. + r_n (q_{n+1} \bar{q}_{n} + q_{n} \bar{q}_{n+1}) \right] (1 + |r_n|^2).
\end{eqnarray}
By taking the asymptotic ansatz
$$
\left\{ \begin{array}{l}
q_n(t) = e^{2it} \left[ \epsilon Q(\epsilon n, \epsilon^2 t) + \mathcal{O}(\epsilon^3) \right], \\
r_n(t) = e^{2it} \left[ \epsilon R(\epsilon n, \epsilon^2 t) + \mathcal{O}(\epsilon^3) \right],
\end{array} \right.
$$
we obtain the system of coupled nonlinear Schr\"{o}dinger equations
at the leading order of $\mathcal{O}(\epsilon^3)$:
\begin{eqnarray}
\label{coupled-1}
i Q_{\tau} + Q_{\xi \xi} + 2 |Q|^2 Q + 4 |R|^2 Q + 2 R^2 \bar{Q} & = & 0, \\
\label{coupled-2}
i R_{\tau} + R_{\xi \xi} + 2 |R|^2 R + 4 |Q|^2 R + 2 Q^2 \bar{R} & = & 0,
\end{eqnarray}
where $\xi = \epsilon n$ and $\tau = \epsilon^2 t$. We can show again that the system of coupled equations
(\ref{coupled-1})--(\ref{coupled-2}) can be reduced to an uncoupled system. By letting $U := Q + R$,
$V := Q - R$ and adding and subtracting equations (\ref{coupled-1}) and (\ref{coupled-2}),
we obtain the two uncoupled nonlinear Schr\"{o}dinger equations:
\begin{eqnarray}
\label{coupled-3}
i U_{\tau} + U_{\xi \xi} + 2 |U|^2 U & = & 0, \\
\label{coupled-4}
i V_{\tau} + V_{\xi \xi} + 2 |V|^2 V & = & 0.
\end{eqnarray}
Hence, the new semi-discrete system (\ref{Sylvlattice1})--(\ref{Sylvlattice2}) is another integrable
semi-discretization of the coupled nonlinear Schr\"{o}dinger equations.

\section{Conclusion}

We have derived a new integrable system of discrete coupled nonlinear Schr\"{o}dinger equations by
considering quadratic and quartic polynomials in the spectral parameter
for the discrete Lax operators satisfying the compatibility condition.
The novel system shares many properties with the integrable Ablowitz--Ladik lattice \cite{AL,AL2}
and triangular--lattice ribbon studied by Vakhnenko \cite{V1,V2}. It has two continuum reductions which are equivalent to uncoupled
modified Korteweg--de Vries and nonlinear Schr\"{o}dinger equations.

The original goal of our study, finding an integrable semi-discretization of the massive Thirring model (\ref{MTM}),
has not been reached in our search. Modifications of the quadratic and quartic polynomials
in the discrete Lax operators did not produce new integrable system of lattice differential equations.
Although the integrable semi-discretizations of the derivative nonlinear Schr\"{o}dinger
equations (\ref{dnls}) and the massive Thirring model in characteristic coordinates have been constructed
in the literature \cite{Tsuchida}, it still remains an open problem to construct
an integrable semi-discretization of the massive Thirring model in physical coordinates.

\vspace{0.25cm}

{\bf Acknowledgement.} We thank P.G. Kevrekidis (University of Massachusetts) 
for suggesting a search for integrable semi-discretization of the MTM system
and Th. Ioannidou (University of Thessaloniki) for collaborating on a search for new semi-discretization of the derivative nonlinear Schr\"{o}dinger
equation. S.A. Bronsard is supported by the NSERC USRA grant for an undergraduate research. D.E. Pelinovsky is supported by
the NSERC Discovery grant.

\end{document}